\documentclass[twocolumn]{el-author}

\usepackage{amsmath}
\usepackage{algorithm}
\usepackage[noend]{algpseudocode}
\usepackage{array}
\usepackage{multirow}
\usepackage{epstopdf} 

\newcommand{\ua}{\uparrow}
\newcommand{\nc}{\newcommand}
\nc{\da}{\downarrow} \nc{\hc}{\hat{c}} \nc{\hS}{\hat{S}}
\nc{\bra}{\langle} \nc{\ket}{\rangle} \nc{\eq}{equation (\ref}
\nc{\h}{\hat} \nc{\hT}{\h{T}}\nc{\be}{\begin{eqnarray}}
\nc{\ee}{\end{eqnarray}}\nc{\rd}{\textrm{d}}\nc{\e}{eqnarray}\nc{\hR}{\hat{R}}\nc{\Tr}{\mathrm{Tr}}
\nc{\tS}{\tilde{S}}\nc{\tr}{\mathrm{tr}}\nc{\8}{\infty}\nc{\lgs}{\bra\ua,\phi|}\nc{\rgs}{|\ua,\phi\ket}
\nc{\hU}{\hat{U}}\nc{\lfs}{\bra\phi|}\nc{\rfs}{|\phi\ket}\nc{\hZ}{\hat{Z}}\nc{\hd}{\hat{d}}\nc{\mD}{\mathcal{D}}
\nc{\bd}{\bar{d}}\nc{\bc}{\bar{c}}\nc{\mc}{\mathcal}\nc{\ea}{eqnarray}\nc{\mG}{\mathcal{G}}\nc{\bce}{\begin{center}}
\nc{\ece}{\end{center}}
\date{12th December 2011}

\begin{document}

\title{Least reliable messages based early termination method for LT soft decoder}

\author{C. Albayrak, C. Simsek and K. Turk}

\abstract{In this paper, we propose a new early termination method (ETM) for Luby transform (LT) belief propagation (BP) decoder. The proposed ETM, which we call least reliable messages (LRM), observes only sign alterations of a small cluster in log-likelihood ratio (LLR) messages passing between nodes in BP decoder. Simulation results and complexity analyzes show that LRM significantly lower computational complexity of early termination section in decoder without any performance degradation and decreases the average decoding iteration amounts compared to conventional ETMs in literature. The method can be easily applied to code families which can be decoded by BP such as low density parity check (LDPC) codes, polar codes and Raptor codes.}

\maketitle

\section{Introduction}
Due to their capacity-approaching and unique rateless properties, there has been a particular interest in using  Luby transform (LT) and Raptor codes, which are members of rateless codes family, over noisy channels \cite{FixedRate,LLR-ADM}. Message-passing algorithms such as belief propagation (BP) are used for decoding of rateless codes. BP iterative decoder uses a pre-set fixed iteration number in order to stop decoding. However, BP mostly converges to original data at an early stage of decoding. Since the decoding continues up to pre-set fixed iteration number, decoder performs redundant processes which cause high computational complexity, decoding latency and energy dissipation. To avoid the aforementioned negations, decoder should be supported by an early termination mechanism to detect convergence and stop decoding.

In literature, there are some early termination methods (ETMs) based on check-sum satisfaction ratio (CSR) for rateless codes \cite{LTstopping1}-\cite{LTstopping4}. CSR is a common success criterion for BP decoding algorithm to observe whether message estimation satisfies constraints imposed by check nodes. Iterative BP decoding algorithm is performed through log-likelihood ratio (LLR) message-passing between nodes. At the end of each iteration CSR decides output bits, re-encodes them and compare with input bits to determine successful convergence.
In this letter we propose a completely new ETM for LT BP decoder which we denote as least reliable messages (LRM) ETM. Our method observes only sign alterations of a small cluster in passing LLR messages between BP nodes.
Results show that proposed LRM ETM significantly reduces the computational complexity of early termination section in decoder without any performance loss and also decreases the average iteration amounts compared to CSR.

\section{BP Decoder for LT Codes}
Tanner graph representation of LT codes contains two types of nodes, check-node (CN) and variable-node (VN). BP decoding algorithm is performed through LLR message-passing between these CNs and VNs iteratively. After running LT decoder for a pre-set fixed iteration amount, decision process is done and decoding is completed \cite{FixedRate,LLR-ADM}. The updating equations of CN and VN in LT BP decoder are given as $m_{c \to v}^{(l)}=sign\left(m_{c}\prod_{v' \neq v}m_{v' \to c}^{(l)} \right)
\times \ 2tanh^{-1} \left[tanh\left({|m_{c}|}/{2} \right)  \prod_{v' \neq v} tanh\left({|m_{v' \to c}^{(l)}|}/{2} \right) \right]$ and $m_{v \to c}^{(l+1)}=\sum_{c' \neq c}m_{c' \to v}^{(l)}$, respectively. Here, $m_{c}$ stands for LLR values of the codewords come from channel and is directly sent to corresponding CN $c$, $m_{c \to v}$ and $m_{v \to c}$ is the outgoing LLR messages from the CN $c$ to VN $v$ and vice versa. $tanh$ and $tanh^{-1}$ represent hyperbolic tangent and its inverse operations, respectively.  Superscript $l$ denotes iteration index. Hard-decision process of BP is given as $m_{v}=m_{c \to v}^{(l)}+m_{v \to c}^{(l+1)}$ and  $\hat{m}_{v}=1$ if $m_{v} \geq 0$, $\hat{m}_{v}=0$, if $m_{v} < 0$. Here, $\hat{m}_{v}$ represents hard value for corresponding VN $v$.

\section{CSR Early Termination Method}
A common criterion for early termination of rateless decoding is observing if the estimated messages $\hat{m}_{v}$ satisfy the constraints imposed by CNs \cite{LTstopping1,LTstopping2}. The criterion controls whether the equation $\hat{m}_{c}\oplus\ \left(\bigoplus_{v}\hat{m}_{v}\right)$ is equal to zero for all CNs, where $\hat{m}_{c}$ stands for hard decision of $m_{c}$ messages, $\oplus$ represents modulo-2 addition and $\bigoplus$ denotes the summation operator for modulo-2 addition. Parenthetical expression represents re-encoding process and rest of it represents compare process. After that, CSR test is calculated by $\mu_{CSR}=s^{(l)}/N_{CN}$, where $s^{(l)}$ is number of satisfied CNs at decoding iteration $l$ and $N_{CN}$ is total number of CNs. The test is satisfied when inequality $\mu_{CSR} \geq \Gamma_{CSR}$ is correct, where $\Gamma_{CSR}$ is a user-defined threshold. This method is known as CSR ETM. LT BP decoder with CSR is presented in Algorithm \ref{alg:CSR}. 
\begin{algorithm}
\caption{LT BP decoder with CSR method:}\label{alg:CSR}
\begin{algorithmic}[1]
\State Calculate $m_{c}$;
\State Set $m_{c \to v}^{(0)}$ and $m_{v \to c}^{(0)}$ messages to zero, $l=0$;
\While{$(l < max\_iter)$ \textbf{and} $(\Gamma_{LC} \ is \ not \ satisfied)$}
\State CN update();	
\State VN update();	
\State Decision();		
\State Calculate CSR and $\Delta$CSR;
\State{$l=l+1;$} 				
\EndWhile
\end{algorithmic}
\end{algorithm}
In the algorithm, the difference between CSR values of two consecutive iterations denoted as $\Delta$CSR. If $\Delta$CSR has a value of ''$0$'' for $\Gamma_{LC}$ amount of consecutive iterations, decoding is terminated \cite{LTstopping3}. $\Gamma_{LC}$ is a user-defined integer value.

\section{Proposed LRM Early Termination Method}
LRM ETM is based on observing sign alterations of a small cluster in $m_{v \to c}$ messages. Since the sign parts of the LLR values are utilized for hard-decision in the decision part of BP, observing sign alterations of $m_v$ during successive iterations can be used to determine whether estimated data bits change. If the estimated data bits stop changing for a number of consecutive iterations ($\Gamma_{LC}$) it can be assumed that decoder successfully converged. To be able to get lowest average iteration amounts, $\Gamma_{LC}$ value should be as low as possible. 

Instead of $m_v$ messages, our proposed method observes sign alterations of $m_{v \to c}$ messages that specify $m_v$. Therefore, our method doesn't require performing ''Decision()'' at each decoding iteration.
On the other hand, proposed LRM method is basically based on the fact that $m_{v \to c}$ messages with lower absolute LLR values are less reliable among entire $m_{v \to c}$ messages \cite{LLR-ADM} and they converge later than messages that have higher absolute LLR values. Therefore, we observe only LRM which is a small cluster of LLR values to determine successful convergence. This simplification also reduces the computational complexity of ETM section significantly. Determination of LRM which means finding the smallest absolute LLR values in all $m_{v \to c}$ messages, can be easily done by using a selection algorithm. We use quickselect algorithm which has low computational complexity \cite{Quicksort}. 

LRM ETM determines LRM to observe sign alterations after running decoder for a few iterations. This is because LT BP decoder typically needs a few iterations to propagate initial channel LLR values.
We call these threshold for iteration numbers as determination condition of LRM (DC-LRM). It is easy to see that larger DC-LRM value increases probability of choosing accurate LRM because better propagation occurs when iteration number increases. On the other hand, DC-LRM shouldn't be larger than minimum iteration number that decoder converged to keep average iteration number as low as possible. 
DC-LRM values are chosen as 45, 28, 22, 18 and 15 for 0.5, 1.0, 1.5, 2.0 and 2.5dB according to simulations, respectively. DC-LRM values for different systems can be determined by simulations and previously loaded to a look-up table. LT BP decoding process with proposed LRM method is presented in Algorithm \ref{alg:LRM}.
\begin{algorithm}
\caption{LT BP decoder with LRM method:}\label{alg:LRM}
\begin{algorithmic}[1]
\State Calculate $m_{c}$;
\State Set $m_{c \to v}^{(0)}$ and $m_{v \to c}^{(0)}$ messages to zero, $l=0$;
\While{$(l < max\_iter)$ \textbf{and} $(\Gamma_{LC} \ is \ not \ satisfied)$}
\State CN update();	
\State VN update();	
\If{$(l ==$ DC-LRM$)$} Quickselect(); \EndIf
\If{$(l >$ DC-LRM$)$} Count sign changes in LRM; \EndIf
\State{$l=l+1;$} 
\EndWhile
\State Decision();
\end{algorithmic}
\end{algorithm}

\section{Complexity Analyzes}
In this section, we analyze the computational complexities of CSR ETM and proposed LRM ETM. We count up computational complexities of considered ETMs and illustrate the results in Table \ref{tab:complexity}.  We assume $abs$, $sign$ and $XOR$ operations have same complexities to simplify the comparison.
\begin{table}[ht]
\begin{center}
\caption{Complexities of ETMs for single iteration}
\label{tab:complexity}
\begin{tabular}{|c|>{\centering\arraybackslash}m{0.85in}|>{\centering\arraybackslash}m{1.04in}|}
\hline
\textbf{Operation} 	      	& \textbf{CSR}												& \textbf{LRM}\\\hline 
$Addition$ 	      			& $N+K(1-\lambda_{1})$										& $N_{B}$\\\hline 
$abs$, $sign$, $XOR$ 	& $N\sum_{d_c=1}^{d_{c_{max}}}d_c \rho_{d_c}$ 			& $N_{B}$\\\hline 
$Compare$ 				& $K$ 														& $N_{B}+2N_{m_{v \to c}}/l_{avg}$\\\hline 
\end{tabular}
\end{center}
\end{table}
In the table, $N$ is coded packet length, $K$ is uncoded packet length, $\lambda_{1}$ is the fraction of VNs of degree $1$, $d_{c}$ is CN degree, $\rho_{d_c}$ is the fraction of CNs of degree $d_c$ and $d_{c_{max}}$ is maximum CN degree. $N_{B}$ symbolizes number of LRM determined by $N_{B}=B*N_{m_{v \to c}}$, where $B$ is the percentage to determine the amount of LRM, $N_{m_{v \to c}}$ is number of all $m_{v \to c}$ messages and calculated by $N_{m_{v \to c}}=N\Omega'(1)$, where $\Omega'(1)$ is average degree of degree distribution chosen for LT code \cite{BSMC}. As we mentioned above, LRM method performs quickselect algorithm only one time for whole decoding process to determine least reliable messages. The quickselect uses less than $2N_{m_{v \to c}}$ compare operations to find the smallest $N_{B}$ items of an array with length $N_{m_{v \to c}}$ \cite{Quicksort}. We add the average effect of quickselect to computational complexities for each iteration by $2N_{m_{v \to c}}/l_{avg}$ comparisons in the table. Here, $l_{avg}$ is average iteration number. 

It should be also emphasized that all operations required for CSR method are performed in every decoding iteration until decoding is terminated, while the operations for LRM method start after decoder runs DC-LRM iterations which does not emphasised in Table \ref{tab:complexity}.

\section{Numerical results}
In this section, we evaluate the bit error rate (BER) performances of LT BP decoding algorithm with and without ETMs over binary-input additive white Gaussian noise (BIAWGN) channel by simulation works. Also, computational complexities of ETMs and average iteration amounts of BP algorithm with LRM and CSR ETMs are compared. For all simulation works and complexity analyzes, we consider the following degree distribution, $\Omega(x) = 0.008x +0.494x^{2}+0.166x^{3}+0.073x^{4}+0.083x^{5}+0.056x^{8}$ $+0.037x^{9}+0.056x^{19}+ 0.025x^{65}+0.003x^{66}$ \cite{BSMC}, code rate of $1/2$, data packet length of $4000$ and fixed iteration number of $100$. 
\begin{figure}[ht]
\centering{\includegraphics[scale=0.29]{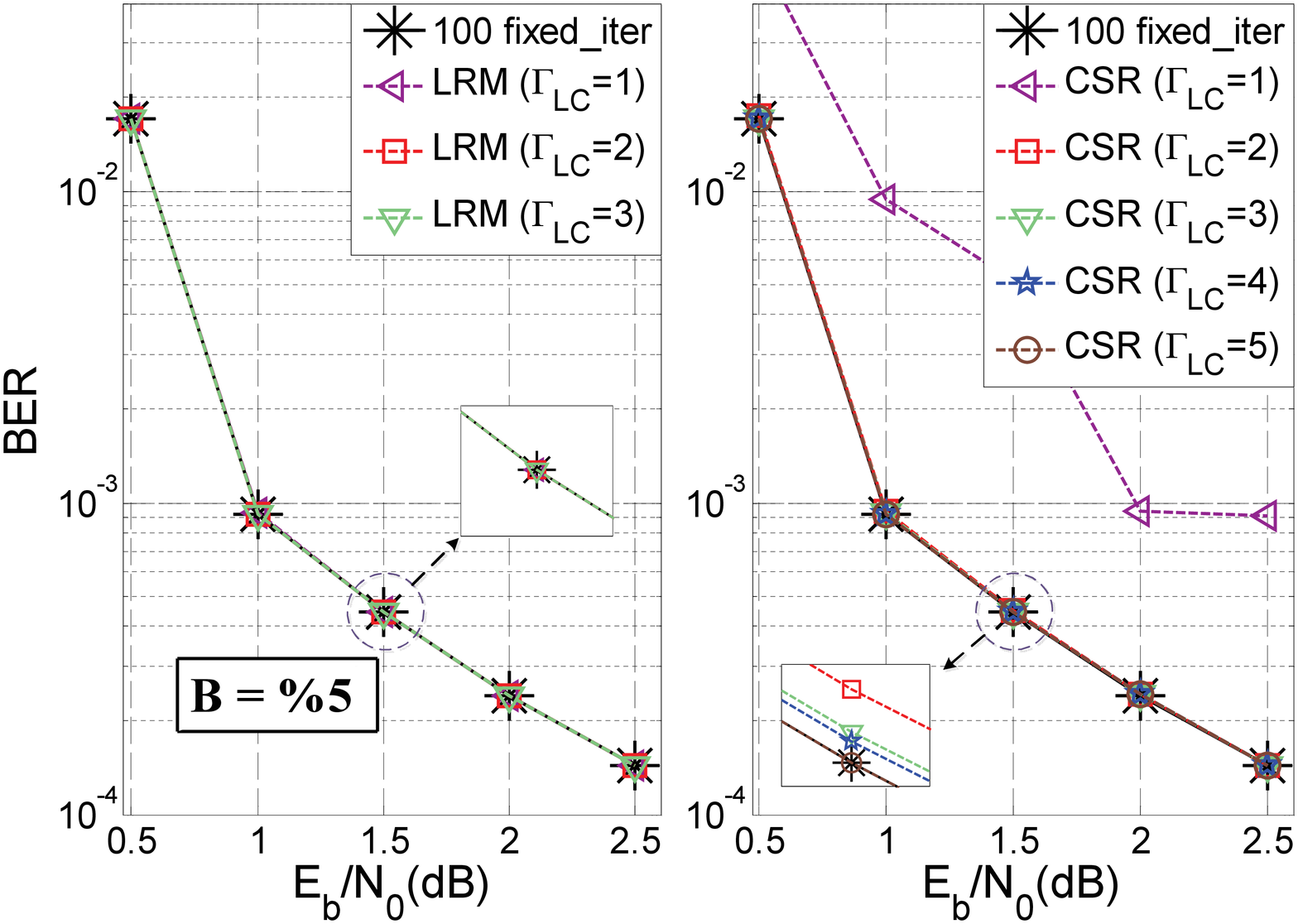}}
\caption{BER curves of LT BP decoder with and without ETMs
\source{}}
\label{fig:snrber}
\end{figure}

Fig. \ref{fig:snrber} illustrates BER curves of LT BP decoder with CSR and proposed LRM ETMs. Simulations are performed for various $N_B$ and $\Gamma_{LC}$ values. Since larger $\Gamma_{LC}$ cause larger average iteration amount, $B=\%5$ is chosen to make $\Gamma_{LC}$ value as small as possible and only the results for $B=\%5$ are illustrated for various $\Gamma_{LC}$. We also provide BER curve for LT BP with 100 fixed iteration number without ETM as a benchmark. This benchmark shows the best BER values that decoder can reach. Differences between benchmark and other BER values indicate that ETMs stop decoding before decoder converges. An ETM shouldn't cause BER performance degradation. As it can be seen in the figure, LRM method with ($\Gamma_{LC}=1$ and $B=\%5$) and CSR with $\Gamma_{LC}=5$ don't cause BER performance degradation. Therefore, performance comparison between ETMs is made with same parameters above.

Table \ref{tab:complexity_iter} compares average iteration amounts of LT BP decoder with selected LRM and CSR methods. Second column in Table \ref{tab:complexity_iter} called ''Decoder Convergence'' is considered as benchmark. LRM ETM has smaller average iteration amounts than CSR but it has slightly higher than benchmark values.
\begin{table}[ht]
\centering
\caption[caption]{Average iteration amounts of LT BP decoder with ETMs and\\\hspace{\textwidth}LT BP decoder successfully converged}
\label{tab:complexity_iter}
\begin{tabular}{|c|>{\centering\arraybackslash}m{0.7in}|>{\centering\arraybackslash}m{0.55in}|>{\centering\arraybackslash}m{0.55in}|}
\hline
\multirow{2}{*}{\boldmath{$E_b/N_0 (dB)$}} & \multirow{2}{*}{\textbf{\begin{tabular}[c]{@{}c@{}}Decoder \\ Convergence\end{tabular}}} & \multicolumn{2}{c|}{\textbf{ETM}} \\ \cline{3-4} 
&	& \textbf{CSR}    & \textbf{LRM}    \\ \hline
$0.5$		& $90.74$ 		& $91.65$		& $91.53$	\\ \hline
$1.0$           & $41.25$		& $45.19$		& $43.73$	\\ \hline
$1.5$           & $28.65$		& $32.45$		& $30.15$	\\ \hline
$2.0$           & $22.84$		& $26.70$		& $24.04$	\\ \hline
$2.5$           & $19.42$		& $23.33$		& $20.37$	\\ \hline
\end{tabular}
\end{table}

Average computation times of ETMs for decoding a code block are compared in Table \ref{tab:complexity_ET} with considered simulation parameters (CSR with $\Gamma_{LC}=5 $ and LRM with $\Gamma_{LC}=1, B=\%5  $). Results show that required computation time of LRM method is significantly lower than CSR. Note that timing results demonstrate only ETM section of decoding process. Furthermore, decoder with proposed LRM method has small average iteration amounts compared to decoder with CSR as shown in Table \ref{tab:complexity_iter}. This provides additional reduction in computation time of whole decoding process.
\begin{table}[ht]
\centering
\caption{Average computation time of EMTs for decoding a code block}
\label{tab:complexity_ET}
\begin{tabular}{|>{\centering\arraybackslash}m{0.7in}|>{\centering\arraybackslash}m{0.5in}|>{\centering\arraybackslash}m{0.5in}|>{\centering\arraybackslash}m{0.75in}|}
\hline
\boldmath{$E_b/N_0 (dB)$} 	& \textbf{CSR(ms)} 	& \textbf{LRM(ms)}		& \textbf{Reduction (\%)} \\ \hline
$0.5$		& $86.18$		& $6.91$		& $91.98$		\\ \hline
$1.0$		& $38.99$		& $3.07$		& $92.13$		\\ \hline
$1.5$		& $26.93$		& $2.01$		& $92.54$		\\ \hline
$2.0$		& $21.80$		& $1.72$		& $92.11$		\\ \hline
$2.5$		& $19.28$		& $1.58$		& $91.80$		\\ \hline
\end{tabular}
\end{table}


\section{Conclusion}
In this paper, we developed a new early termination method for LT BP decoder to avoid redundant processes which cause high computational complexity, decoding latency and energy dissipation.
Simulation results and complexity analyzes show that proposed LRM method significantly lower complexity and computation time of early termination section in decoder without BER performance degradation and decreases the average iteration amounts compared to conventional CSR ETM. The method can be easily applied to code families which can be decoded by BP such as low density parity check (LDPC) codes, polar codes and Raptor codes. The best way to compare ETMs can be done by hardware implementation which will be held in future.

\vskip5pt

\noindent C. Albayrak, C. Simsek and K. Turk (\textit{Department of Electrical Electronics Engineering, Karadeniz Technical University, Trabzon 61080, TR})

\vskip3pt

\noindent E-mail: kadir@ktu.edu.tr


\begin{thebibliography}{}

\bibitem{FixedRate} 
Sivasubramanian, B., Leib, H.: 'Fixed-rate Raptor codes over rician fading channels', \textit{IEEE Trans. Vech. Tech.}, 2008, 57, (6), pp. 3905-3911, doi: 10.1109/TVT.2008.923664

\bibitem{LLR-ADM}
Turk, K., Fan, P.: 'Adaptive demodulation using rateless codes based on maximum a posteriori probability', \textit{IEEE Commun. Lett.}, 2012, 16, (8), pp. 1284-1287, doi: 10.1109/LCOMM.2012.060112.120772

\bibitem{LTstopping1}
AbdulHussein, A. , Oka, A., Lampe, L.: 'Decoding with early termination for rateless (Luby transform) codes', IEEE Wireless Commun. and Networ. Conf. (WCNC), Las Vegas, USA, April 2008, pp. 249-254, doi 10.1109/WCNC.2008.49

\bibitem{LTstopping2}
AbdulHussein, A., Oka, A., Lampe, L.: 'Decoding with early termination for Raptor codes', \textit{IEEE Commun. Lett.}, 2008, 12, (6), pp. 444-446, doi: 10.1109/LCOMM.2008.080260

\bibitem{LTstopping3}
Orozco, V.L., Yousefi, S.: 'Trapping sets of fountain codes', \textit{IEEE Commun. Lett.}, 2010, 14, (8), pp. 755-757, doi: 10.1109/LCOMM.2010.08.100548

\bibitem{LTstopping4}
Chen, Y.-M., Lee, H.-C., Ueng, Y.-L., Yeh, C.-Y.: 'Flooding-assisted informed dynamic scheduling for rateless codes', IEEE Wireless Commun. and Networ. Conf. (WCNC), Paris, France, April 2012, pp. 173-177, doi 10.1109/WCNC.2012.6214065

\bibitem{Quicksort}
Sedgewick, R., Wayne, K.: 'Algorithms' (Princeton University, Addison-Wesley, USA, 2011, 4th edition)

\bibitem{BSMC}
Etesami, O., Shokrollahi, A.: 'Raptor codes on binary memoryless symmetric channels', \textit{IEEE Trans. Inf. Theory}, 2006, 52, (5), pp. 2033-2051, doi: 10.1109/TIT.2006.872855

\end{thebibliography}
\end{document}